\documentclass[prd,aps,eqsecnum,amsmath,floatfix,nofootinbib,preprint,tightenlines]{revtex4-2}

\usepackage{latexsym}
\usepackage{graphicx}
\usepackage[dvipsnames]{xcolor}

\usepackage{hyperref}

\def\bibi{\bibitem}

\def\d{\delta}
\def\g{\gamma}

\def\j{\psi}

\def\m{\mu}
\def\n{\nu}

\def\p{\pi}                     
\def\th{\theta}                  
\def\r{\rho}                    

\def\J{\Psi}

\def\S{\Sigma}
\def\U{\Upsilon}

\def\ca{{\cal A}}

\def\cd{{\cal D}}

\def\cl{{\cal L}}

\def\car{{\cal R}}

\def\cv{{\cal V}}

\def\cbo{{\,\raise-.15ex\Sc [\,}}                       

\def\svev#1{\left\langle #1\right\rangle}       

\def\ddt#1{{\buildrel {\hbox{\LARGE .\kern-2pt.}} \over {#1}}}

\def\leqx{\,\raisebox{-1.0ex}{$\stackrel{\textstyle <}{\sim}$}\,}

\def\tr{{\rm tr}\,}

\def\half{{1\over 2}}
\def\Re{{\rm Re\,}}

\long \def \blockcomment #1\endcomment{}

\def\ttl#1{{\it #1}}

\def\bj{\overline\psi}
\def\bJ{\overline\Psi}

\def\textit#1{{\it \!\!\! #1 \!\!}}

\def\SU{{\rm SU}}
\def\U{{\rm U}}

\def\textvev#1{\langle #1 \rangle}

\begin{document}

\begin{center}
{\large{\bf Conserved currents in five-dimensional proposals\\[2mm]
            for lattice chiral gauge theories }}\\[8mm]
Maarten Golterman$^a$ and Yigal Shamir$^b$\\[8 mm]
$^a$Department of Physics and Astronomy, San Francisco State University,\\
San Francisco, CA 94132, USA\\
$^b$Raymond and Beverly Sackler School of Physics and Astronomy,\\
Tel~Aviv University, 69978, Tel~Aviv, Israel\\[10mm]
\end{center}

\begin{quotation}
We apply the Grabowska--Kaplan framework \cite{GK},
originally proposed for lattice chiral gauge theories, to QCD.  We show that
the resulting theory contains a conserved and gauge invariant
singlet axial current, both on the lattice and in the continuum limit.
This must give rise to a difference with QCD, with the simplest possibility
being a superfluous Nambu--Goldstone boson in the physical spectrum
not present in QCD.  We find a similar unwanted conserved current in the recent
``disk'' formalism \cite{Kd,KS}, this time limiting ourselves to
the continuum formulation.  A similar problem is expected
when either of these formalisms is used for its original goal
of constructing lattice chiral gauge theories.  Finally we discuss
a conjecture about the possible dynamics that might be associated
with the unwanted conserved current, and the fate of 't~Hooft vertices.
\end{quotation}

\newpage
\section{\label{intro} Introduction}
The original goal of Kaplan's domain-wall fermions (DWFs) was to construct
lattice chiral gauge theories \cite{KDWF}.  Much later, a concrete realization
of the original idea was proposed by Grabowska and Kaplan \cite{GK}.
As in the standard lattice formulation of QCD using DWFs \cite{YSdwf,FS},
the basic geometry can be taken to be a five-dimensional ``slab,''
with Weyl fermions of opposite chiralities residing on the two
four-dimensional boundaries.

In QCD, the four-dimensional gauge field is taken to be independent of
the fifth coordinate.  The Weyl fermions on both boundaries (or ``walls'')
couple to the gauge field with equal strength, thereby forming Dirac fermions:
one Dirac fermion per each five-dimensional DWF field.
By contrast, in the Grabowska--Kaplan framework the gauge field
inside the five-dimensional (5D) slab is defined via gradient flow \cite{MLGF}
in the fifth direction, with the dynamical four-dimensional (4D) gauge field
on the near wall serving to start the flow.
The goal is that, while preserving gauge invariance,
the gauge field will die out well before reaching the Weyl fermion on
the far wall, thereby decoupling it from the gauge field.
If successful, then only the 4D Weyl fermion on the near wall
of each 5D slab would remain coupled to the gauge field.
All these Weyl fermions can be chosen to have the same handedness,
and the construction could thus be used for a non-perturbative, gauge invariant
definition of chiral gauge theories in four dimensions.

Recently, another proposal was put forward by Kaplan \cite{Kd} and by
Kaplan and Sen \cite{KS}.  In the new proposal the slab geometry
is replaced by a disk geometry, and the rim of the disk
is identified with one of the four physical dimensions.\footnote{%
The remaining three physical dimensions are cartesian.
While we focus on $4+1$ dimensions in this paper, the discussion
generalizes straightforwardly to $2+1$ dimensions.
}
Remarkably, the rim supports a single Weyl fermion of one chirality only
\cite{Kd}.  Moreover, this feature appears to survive lattice discretization
\cite{KS}, thereby circumventing the no-go theorems \cite{KSmit,NN}.
By construction, the dynamical gauge field resides on the rim of the disk,
and is extended into the whole disk via a radial gradient flow,
once again maintaining gauge invariance.

In both the slab and disk geometries,
it was argued that a necessary condition that the degrees of freedom
in the extra dimension will fully decouple in the infrared
is that the (chiral) fermion spectrum of the target 4D gauge theory
satisfy the anomaly cancellation condition for the gauge symmetry.
By contrast, if the fermion spectrum suffers from a gauge anomaly,
then the effective low-energy 4D theory will remain non-local.
Thus, both formalisms pass an important consistency test.

The question arises whether there are other possible stumbling blocks
for the successful construction of lattice chiral gauge theories.
In order to examine this question,
here we turn our attention to the global flavor symmetries
and their associated conserved currents.
Our main finding is that, in both the slab and disk geometries,
there is always one ``superfluous'' 4D current
which is both gauge invariant and conserved,
and which is not present in the target gauge theory.
In order to highlight the persistence of this problem,
we examine the application of the slab and disk frameworks
for an alternative lattice definition of vector-like theories,
taking one-flavor QCD as the target 4D theory for our main example.
In the case of a vector-like theory, the unwanted conserved current
is the singlet axial current, which should be anomalous.
In the case that the target gauge theory is chiral, the additional
conserved current is the total fermion number,
which, once again, should be anomalous.

In Sec.~\ref{QCD} we consider the slab geometry, focusing mainly on
one-flavor QCD to demonstrate the issue.  We work on the lattice,
and thus our conclusions apply both at finite lattice spacing
and in the continuum limit.  In Sec.~\ref{disk} we discuss the novel disk
framework, limiting ourselves to the continuum formulation.  Our findings
are similar to those of the slab framework.  In Sec.~\ref{tHooft}
we conjecture on the dynamics that might be associated with
the superfluous conserved axial current, focusing on the fate of
the 't~Hooft vertices of the target 4D theory.  We conclude in Sec.~\ref{conc}.
In the appendix we discuss lattice gradient flows.

\section{\label{QCD} Conserved currents in the Grabowska--Kaplan framework}
We discuss here the application of the Grabowska--Kaplan (GK) formalism
\cite{GK} to QCD, using the one-flavor theory as an example.
The main conclusion is that the GK framework gives rise to
a singlet axial current which is simultaneously gauge invariant and conserved.
The straightforward interpretation is that this, in turn, gives rise
to a superfluous Nambu-Goldstone boson (NGB) in the physical spectrum.

We begin with a lattice setup containing two DWFs.
In the standard DWF formulation of QCD \cite{YSdwf,FS}, this provides for
the fermion content of the two-flavor theory.  However,
within the GK framework only the Weyl fermion on the near wall
of each DWF is expected to couple to the gauge field.
This will leave us with a total of two Weyl fermions (of opposite chiralities),
in agreement with the fermion content of the one-flavor theory.

In more detail, the lattice theory contains two five-dimensional GK fields,
both in the fundamental representation of SU(3).
The fifth coordinate takes values $s=0,1,\ldots,N_5-1$.
We assume that one of the GK fields, denoted $\J^{(R)}$,
has a right-handed (RH) Weyl field on the near wall,
while the other, denoted $\J^{(L)}$,
has a left-handed (LH) Weyl field on the near wall.
Taken together, we thus have a single Dirac fermion,
the matter content of one-flavor QCD.
The (bare) quark fields are identified with the 5D fields on the $s=0$ layer,
namely, $\j_R(x)=\J^{(R)}(x,s=0)$ and $\j_L(x)=\J^{(L)}(x,s=0)$.
For simplicity, we assume that the quark mass is zero.

In practice, flipping the chirality of the Weyl fermion on the near wall
is done by flipping the sign of $\g_5$
everywhere in the lattice action.\footnote{
In the standard DWF formulation of QCD, this operation can be undone
by a reflection in the fifth dimension, because the gauge field
is independent of the fifth coordinate.}
Therefore, there is no continuous symmetry that interchanges
the two 5D fields $\J^{(R)}$ and $\J^{(L)}$.
However, each 5D field is endowed with an exact U(1) symmetry
that acts on that field only.
Following closely Refs.~\cite{FS,YSan},
the corresponding Noether currents are
\begin{subequations}
\label{currLR}
\begin{eqnarray}
\label{currR}
\car_\m(x) &=& \half \sum_{s=0}^{N_5-1} \left(
 \bJ^{(R)}_{x,s}(1+\g_\m) U_{x,s,\m} \J^{(R)}_{x+\hat\m,s}
 -\bJ^{(R)}_{x+\hat\m,s}(1-\g_\m) U_{x,s,\m}^\dagger \J^{(R)}_{x,s} \right) ,
\hspace{7ex}
\\
\label{currL}
\cl_\m(x) &=& \half \sum_{s=0}^{N_5-1} \left(
 \bJ^{(L)}_{x,s}(1+\g_\m) U_{x,s,\m} \J^{(L)}_{x+\hat\m,s}
 -\bJ^{(L)}_{x+\hat\m,s}(1-\g_\m) U_{x,s,\m}^\dagger \J^{(L)}_{x,s} \right) .
\end{eqnarray}
\end{subequations}
These currents are both gauge invariant and conserved.
Notice that unlike in the standard DWF formulation of QCD,
here the four-dimensional link variables
depend on the fifth coordinate $s$ via the gradient flow.

We may alternatively construct a vector and an axial current,
\begin{subequations}
\label{currVA}
\begin{eqnarray}
\label{currV}
\cv_{\m} &=& \car_\m+\cl_\m \ ,
\\
\label{currA}
\ca_{\m} &=& \car_\m-\cl_\m \ .
\end{eqnarray}
\end{subequations}
These currents, too, are gauge invariant and conserved.
The $\U(1)$ symmetry associated with the vector current $\cv_\m$
rotates $\j_R$ and $\j_L$ with the same phase.
This is baryon number symmetry $\U(1)_B$, a good symmetry of (one-flavor) QCD.
The other $U(1)$ symmetry, associated with the current $\ca_\m$,
rotates $\j_R$ and $\j_L$ with opposite phases; this is
the axial symmetry $\U(1)_A$, which is anomalous in QCD.
The problem is thus that in the GK framework
the axial symmetry is an exact symmetry, too.
As a result, also the (singlet) axial current $\ca_\m$
is both conserved and gauge invariant within the GK framework.
This generates a conflict between the standard properties of (one-flavor) QCD,
and the features of its GK formulation.
If the GK framework for regulating QCD leads to a consistent
continuum limit, that continuum limit must be different from
the one obtained from any of the standard lattice regularizations of QCD.

Let us elaborate on this conflict.  In any standard formulation of QCD,
the axial current $A_\m$ is anomalous,
\begin{equation}
\label{anax}
\partial_\m A_\m = X \ , \qquad X = cg^2\, \tr(F\tilde{F}) \ ,
\end{equation}
with $c$ some non-vanishing numerical constant.
This gives rise to anomalous Ward--Takahashi identities (WTIs).
Consider for example the momentum-space WTI (in the
continuum and chiral limits)
\begin{equation}
\label{WTIan}
ip_\m \svev{A_\m\, \eta}(p) = \svev{X\, \eta}(p) + \S \ ,
\end{equation}
where $\eta = \bj_L\j_R-\bj_R\j_L$ is the pseudoscalar density,
and $\S = \svev{\bj_L\j_R+\bj_R\j_L}$ is the fermion condensate.
Thanks to the anomalous term, this WTI does not require the existence
of any massless particle when $\S\ne 0$, consistent with the
large mass of the $\eta'$ meson in QCD.

By contrast, within the GK formulation of the massless one-flavor theory,
the gauge invariant current $\ca_\m$ has no anomaly,
and, after taking the continuum limit, we obtain the WTI\footnote{
At finite lattice spacing, $p_\m$ is replaced by $(2/a)\sin(ap_\m/2)$.
}
\begin{equation}
\label{WTIax}
ip_\m \svev{\ca_\m\, \eta}(p) = \S \ .
\end{equation}
Note that now $\S$ is a true order parameter, since the axial symmetry
$\U(1)_A$ is exact in the GK formulation.
The next step is to decompose the correlator in terms of invariant amplitudes,
requiring translation and Lorentz invariance.\footnote{%
  After the analytic continuation to Minkowski space.
}
Thanks to the simple form of the correlator,
it depends on only a single invariant amplitude,
\begin{equation}
\label{iAmp}
\svev{\ca_\m\, \eta} = -i p_\m F(p^2) \ .
\end{equation}
Substituting this back into Eq.~(\ref{WTIax}), the unique solution is
$F(p^2) = \S/p^2$, or equivalently,
\begin{equation}
\label{WTIpole}
\svev{\ca_\m\, \eta} = -i \S\, \frac{p_\m}{p^2} \ .
\end{equation}
Provided that chiral symmetry breaking takes place and $\S\ne 0$,
this result exhibits the pole of the Nambu--Goldstone boson.
This new NGB, being a singlet pseudoscalar meson,
would signal a breakdown of universality in QCD.
This is our main conclusion.

As discussed in Ref.~\cite{HK}, it is possible
to split the current $\ca_\m$ into several pieces,
one of which will behave as the---anomalous---axial current of QCD.
Nonetheless, the existence within the GK framework
of the axial current $\ca_\m$, which is both gauge invariant and conserved,
means that there is no escape from the WTI~(\ref{WTIax}),
and its consequences for the physics of the theory.

The existence of a singlet pseudoscalar NGB in the physical spectrum,
which is absent from the standard formulation of QCD,
is a completely general phenomenon within the GK formulation.
Generalizing the lattice setup to QCD with $N_f$ flavors,
it is easy to see that the global symmetry of the GK formulation
will be $\U(N_f)_L\times\U(N_f)_R$, and not
$\SU(N_f)_L\times\SU(N_f)_R\times \U(1)_B$ as expected.
Once again, we will have a singlet axial current
which is both gauge invariant and conserved, and not anomalous.
Again, assuming that the theory confines and breaks its chiral symmetry,
and requiring translation and Lorentz invariance,
then, in addition to the expected pions for $N_f\ge 2$,
there will be a superfluous, singlet pseudoscalar NGB.

The problem persists when our goal is to construct a chiral gauge theory.
A chiral gauge theory in four dimensions can be formulated in terms of
LH fields only, and the total fermion number current is then
always anomalous.  But within the GK formulation,
again there is a conserved and gauge invariant current
associated with the total fermion number.

The violation of the global axial charge in vector-like theories,
and of the total fermion number in chiral gauge theories,
is believed to arise from 't~Hooft vertices.
In Sec.~\ref{tHooft} we discuss a conjecture about the fate
of instantons and 't~Hooft vertices within the GK formulation.

\section{\label{disk} Conserved currents in the disk framework}
It is clear from the discussion of the previous section
that the superfluous current, which is simultaneously
conserved and gauge invariant, originates from the existence
of a fermion number symmetry for each 5D field separately.
This implies the existence of a conserved and gauge invariant 5D current
for each GK field, from which one can construct a 4D conserved current.
One linear combination of the 4D conserved currents will always be superfluous,
as explained in the previous section.

In this section we turn to the disk framework, and demonstrate
the existence of a similar, superfluous four-dimensional current.
One can envisage various ways of discretizing the disk framework.
For example, it is fairly obvious that there exist lattice discretizations
that will preserve a discrete subgroup of the rotational symmetry
of the disk. By contrast, in Ref.~\cite{KS}, a discretization based on
a ``trimmed'' regular hypercubic lattice was preferred.
In view of these rather different options for the lattice discretization,
we will limit the discussion in this section to the continuum case only.

\subsection{\label{cdisk} The conserved four-dimensional current}
We first briefly introduce the disk framework.
We will mostly disregard the gauge field,
since it plays no role in the (classical) conservation equation.
We stress, however, that in the presence of the gauge field,
the current is gauge invariant.  Moreover, since continuous symmetries
do not have anomalies in odd spacetime dimensions, the conservation
of the gauge invariant 5D current holds to all orders in perturbation theory.
A similar conservation equation holds
for the 5D lattice discretization of Ref.~\cite{KS} as well.

In the $(x,y)$ plane, the fermions are restricted to a disk of radius $R$.
In addition, there are three cartesian coordinates, denoted $z_i$, $i=1,2,3$.
The 5D fermion field satisfies boundary conditions defined in terms of
the radial projectors $P_\pm^r=\half(1\pm\g_r)$,
where $\g_r=\g_x\cos\theta+\g_y\sin\theta$.  As was shown in Refs.~\cite{Kd,KS},
this construction gives rise to a single Weyl fermion on the rim of the disk.
Hence, the rim of the disk is identified with the fourth ordinary dimension,
which thus has a finite length $L=2\p R$ and periodic boundary conditions.
It is described by the coordinate $R\theta$, with $0\le\theta <2\p$.
The radial direction of the disk, with coordinate $0\le r\le R$,
corresponds to the fifth direction of the slab geometry.\footnote{%
  The limit $R\to\infty$ in the disk geometry corresponds for the slab geometry
  to simultaneously taking the limits $N_5\to\infty$ together with
  the limit $L\to\infty$ for the cartesian direction identified
  with the rim of the disk.}

The conservation equation for the Noether current of the U(1) symmetry
of a given 5D fermion field takes the form
\begin{equation}
\label{div5j}
0 = \sum_i \partial_i j_i + \partial_x j_x + \partial_y j_y
= \sum_i \partial_i j_i + \frac{1}{r}\,\partial_r (r j_r)
+ \frac{1}{r}\,\partial_\th j_\th \ .
\end{equation}
The middle expression is the divergence of the 5D current in
cartesian coordinates, while in the rightmost expression we switched
to radial coordinates for the $(x,y)$ plane containing the disk.
$j_r$ ($j_\th$) is the component of the current
in the radial (tangential) direction.  Like the fermion field itself,
the 5D current is restricted to $r<R$.

We define the 4D current by integrating along rays,
\begin{subequations}
\label{J4def}
\begin{eqnarray}
\label{Ji}
J_i(\th;z_i) &=& \frac{1}{R} \int_0^R rdr\, j_i(r,\th;z_i) \ , \qquad
i=1,2,3 \ ,
\\
\label{Jth}
J_\th(\th;z_i) &=& \rule{0ex}{4ex} \int_0^R dr\, j_\th(r,\th;z_i) \ .
\end{eqnarray}
\end{subequations}
Notice that the integration measure for the three transverse components
is $(r/R)dr$, whereas the factor $r/R$ is absent from
the definition of $J_\th$, the component associated with the
tangential direction along the rim of the disk.
As we will now demonstrate, this is the right choice that
leads to the conservation of the 4D current.
Suppressing the coordinates $(\th;z_i)$ one has
\begin{eqnarray}
\label{cJ4}
\sum_\m \partial_\m J_\m
&\equiv&
\sum_i \partial_i J_i
+ \frac{1}{R}\,\partial_\th J_\th
\\
&=& \frac{1}{R} \int_0^R rdr
\left( \sum_i \partial_i j_i + \frac{1}{r}\,\partial_\th j_\th \right)
\nonumber\\
&=&\rule{0ex}{4ex}
- \frac{1}{R} \int_0^R dr\,\partial_r (r j_r)
\nonumber\\
&=&\rule{0ex}{4ex}
\frac{1}{R} (rj_r)\Big|_{r\to 0} \ -\ \frac{1}{R} (rj_r)\Big|_{r=R} \ =\ 0 \ .
\nonumber
\end{eqnarray}
On the first line, the derivative of the tangential component
is $(1/R)\partial/\partial\th$, since $Rd\th$ is the line element
along the rim of the disk. On the second line we substituted
the definitions~(\ref{J4def}), and on the next line we used
the conservation of the 5D current, Eq.~(\ref{div5j}).

As might be expected, the divergence of the 4D current
ends up being a surface term of the radial integral.
The first term on the last line vanishes trivially,
because of the $r\to 0 $ limit.\footnote{%
  Since the radial direction is ill-defined at the center,
  the lower end of the integration is defined as $r\to 0$,
  rather than $r=0$.
}
In addition, the surface term on the rim of the disk
vanishes identically thanks to the boundary conditions.
For definiteness let us assume that the boundary conditions at $r=R$ are
$P_+^r\j=0$, $\bj P_-^r=0$.\footnote{
  These boundary conditions are consistent with the (Minkowski) relation
  $\bj = \j^\dagger \g_0$.
  We identify $\g_0$ as the Dirac matrix associated with one of the $z_i$
  directions, hence it anticommutes with $\g_r$.
}
It follows that, on the rim, the radial component of the current is
$j_r = \bj \g_r \j = \bj P_+^r \g_r P_-^r \j = 0$.
This completes the proof.

\subsection{\label{annulus} Annulus geometry}
It is interesting to explore the relation between the slab and disk geometries.
The connection is provided by an annulus geometry.  We start from the disk,
and cut out a smaller disk of radius $R'<R$.  For the boundary conditions
we specified above on the outer rim $r=R$, the boundary conditions
on the inner rim $r=R'$ will be $P_-^r\j=0$, $\bj P_+^r=0$.

The annulus geometry is topologically equivalent to the slab geometry.
Starting from the slab geometry, let us take one of the four physical directions
to be finite, and with periodic boundary conditions.
The two-dimensional manifold consisting of this physical direction
together with the (finite) fifth direction is then topologically
equivalent to an annulus.

In the annulus geometry, the definition of the conserved
4D current remains the same as in Eq.~(\ref{J4def}),
except that the lower end of the radial integration is now $r=R'$.
The proof that this current is conserved works
as in Eq.~(\ref{cJ4}), with one notable change.
The divergence of the 4D current is now the difference of two
boundary terms
\begin{equation}
\label{annJ}
\sum_\m \partial_\m J_\m
\ =\ \frac{1}{R} (rj_r)\Big|_{r=R'} \ -\ \frac{1}{R} (rj_r)\Big|_{r=R} = 0 \ ,
\end{equation}
and both terms vanish thanks to the boundary conditions imposed
on the respective boundary.

\section{\label{tHooft} Dynamical considerations: the fate of 't Hooft vertices}
In asymptotically free 4D gauge theories, it is widely believed
that violation of the axial charge in vector-like theories, and of
fermion number in chiral gauge theories, comes from instantons\footnote{%
  Other topologically non-trivial configurations may contribute as well.
}
through the effective 't~Hooft interactions they induce \cite{hooft,SC}.

Let us focus once again on the example of QCD with $N_f$ flavors.
Formulating the theory in terms of LH and RH fields in the fundamental
representation, the Weyl fields are
$\j_{Ri}$ and $\j_{Li}$, where $i=1,2,\ldots,N_f$.
The 't~Hooft interactions induced by the instantons violate the conservation of
the fermion number of each Weyl field individually.
In terms of global symmetries,
the group $\SU(N_f)_L\times\SU(N_f)_R\times \U(1)_B$
is respected by the 't~Hooft interaction, while the global axial charge
is not conserved.

Let us now turn to the GK or disk frameworks.  As we have discussed
in the previous sections, now the individual fermion numbers
are conserved for each 5D field separately.  The theory as a whole,
and its 't~Hooft interactions in particular, must therefore be invariant under
the larger symmetry group $\U(N_f)_L\times\U(N_f)_R$.
We emphasize that this behavior is completely general.  In particular,
it is true regardless of how the 4D gauge field is extended into 5D,
as long as the construction preserves the 4D gauge invariance.
We will argue that, since this symmetry is preserved on the lattice,
it remains true for the effective 4D theory in the continuum limit of the
lattice theory.

We will not attempt to discuss the 5D dynamics in complete generality,
because the details can vary a lot, depending on how the 4D gauge field
is extended into 5D.
Instead, we consider in the appendix a family of lattice gradient flows
suitable for both the slab and disk geometries.  These flows are designed
such that it is expected that all instantons will shrink in size
under the flow, and eventually disappear.
In particular, using such a lattice flow in the GK framework,
we expect that in the limit of an infinite fifth dimension, $N_5\to\infty$,
the flowed gauge field on the far wall will always be
a pure gauge with trivial topology,
and hence that the far-wall Weyl fermions fully decouple.

Returning for simplicity to the example of the one-flavor theory
discussed in Sec.~\ref{QCD}, in the field of an instanton we expect to have
one zero mode for (say) the fermion field $\j_R$, and another one for the
antifermion field $\bj_L$.  The corresponding 't~Hooft interaction is thus,
schematiclaly, $\bj_L(x^0) \j_R(x^0)$, where $x^0_\m$ are
collective coordinates: the coordinates of the center of the instanton.
These zero modes will arise from the Weyl fields that reside on the near walls
in the GK framework, or on the rims of the disk in the alternative framework.
The resulting 't Hooft operator is
\begin{equation}
\label{1fphys}
O_{4D} \sim \bj_L(x^0) \j_R(x^0) \ .
\end{equation}
The notation $O_{4D}$ is to indicate that it accounts for the zero modes
of the 4D fields of the target theory.

By itself, the operator $O_{4D}$ violates the
individual fermion numbers of both of the 5D fields $\J^{(R)}$ and $\J^{(L)}$.
But the GK and disk frameworks preserve these individual fermion numbers,
hence there must exist additional zero modes,
to compensate for the 4D zero modes.

What is the dynamics responsible for the existence of such additional
zero modes?  We conjecture that one way for them to arise is as follows.
As explained above, under the class of lattice flows we introduce in the appendix,
the size of the instanton keeps shrinking with the fifth coordinate.
After a long enough flow, the instanton's size will become comparable to the
lattice spacing $a$, which enables it to eventually disappear altogether.
The 5D point $(x,s)$ inside the bulk where the instanton disappears
must exhibit a {\em dislocation} of the flowed gauge field.
We conjecture that a new zero mode may develop
with support on this dislocation.\footnote{
  For related setups where a bulk zero mode appears,
  see Refs.~\cite{AF,AFK}.
}
In order to restore the conservation of the individual fermion numbers
of the 5D fields, the bulk zero modes must have opposite U(1) charges
from the 4D zero modes.  Unlike the power-law decay
of the familiar instanton zero modes, we expect the bulk zero modes
to be exponentially localized.\footnote{
  Such bulk zero modes would resemble the localized zero modes
  of the supercritical Wilson kernel of DWFs
  \cite{lcl1,lcl2,lcl3}.
}

Let us illustrate the role of the novel bulk zero modes,
again using the example of the one-flavor theory.
Also, for simplicity, we will consider the GK framework,
but a similar reasoning applies to the disk framework as well.
For $N_f=1$ we expect two bulk zero modes, which are
represented by the operator
\begin{equation}
\label{1fbulk}
O_{\rm bulk} \sim \bJ^{(R)}(x,s) \J^{(L)}(x,s) \ .
\end{equation}
The total 't~Hooft vertex is the product of the terms
coming from the near wall and from the bulk,
\begin{equation}
\label{1ftot}
O_{\rm tot} = O_{4D} O_{\rm bulk}
\sim \bj_L(x^0) \j_R(x^0) \bJ^{(R)}(x,s) \J^{(L)}(x,s)  \ .
\end{equation}
Now the individual fermion numbers of the two 5D fields are preserved,
as required by the $\U(N_f)_L\times\U(N_f)_R$ global symmetry.\footnote{%
A similar result was anticipated in Ref.~\cite{GK} for the
case of a continuum flow. See also the appendix.}

We will next argue that, in the continuum limit, $O_{\rm tot}$ vanishes
as an operator acting on the states of the effective 4D theory.
This would imply that the $\U(N_f)_L\times\U(N_f)_R$ symmetry of the
underlying lattice theory is inherited by the effective 4D theory.

We start by examining the expectation value of $O_{\rm tot}$ itself,
which satisfies
\begin{equation}
\label{vevtot}
\svev{O_{\rm tot}} \le C e^{-2Ms} \ ,
\end{equation}
for some constant $C$, which in turn is independent of $s$.
Here $M=O(a^{-1})$ is the mass of the bulk 5D fermions.\footnote{
Also known as the domain-wall height.}
The bound arises because
the propagator in the fifth direction falls off like $e^{-M|s-s'|}$,
and the effective 4D fields are located on the boundary $s'=0$.
As long as the instanton's size is large compared to
the lattice scale,\footnote{
See the appendix for details.}
one expects that the product $Ms$ diverges in the continuum limit.
Hence, the expectation value of $O_{\rm tot}$ vanishes.

The above behavior generalizes to any correlation function
involving $O_{\rm tot}$ together with any number of insertions of 4D fields
residing on the $s'=0$ boundary.  In order to contract the fermion fields
contained in $O_{\rm bulk}$, for each 5D field we will need
one propagator from the position $(x,s)$ of the dislocation to the boundary.
As we have just argued, this propagator is bounded from above by $e^{-Ms}$,
again leading to a bound similar to Eq.~(\ref{vevtot}) for the
correlation function under consideration.

In the above argument we have used that the contraction
$\textvev{\J^{(L)}(x,s) \bJ^{(R)}(x,s)}$ vanishes identically,
because it does not preserve the individual fermion numbers of
the 5D fields.  The same result generalizes to the thermodynamical limit.
In order to take the thermodynamical limit we introduce a small mass term
$m_q (\bj_R \j_L + \bj_L \j_R)$.  Since the mass term couples the two
5D fields, now $\textvev{\J^{(L)}(x,s) \bJ^{(R)}(x,s)}$ is non-zero,
and (quark-disconnected) terms that include this contraction
must be considered as well.
However, any propagation from one 5D field to the other must go through the
mass insertion, which in turn lives on the boundary. Hence\footnote{
The $\leqx$ symbol indicates the presence of a proportionality factor,
similarly to Eq.~(\ref{vevtot}).}
\begin{equation}
\label{JLJRm}
\svev{\J^{(L)}(x,s) \bJ^{(R)}(x,s)} \leqx m_q e^{-2Ms} \ .
\end{equation}
It follows that a uniform upper bound by $e^{-2Ms}$ still applies,
and once again the correlation function vanishes in the continuum limit.
The upshot is that the bulk part of the
new 't Hooft vertex suppresses the original
4D 't Hooft vertex, thus providing a dynamical
understanding of the exact, superfluous $\U(1)_A$ symmetry.

In this section we have illustrated via a concrete scenario
how the familiar 't~Hooft vertices of QCD can get modified.
Ultimately, in general, the key point is that
$\U(1)_A$ is an exact symmetry in a gauge invariant formalism.
Regardless of the details of the 5D dynamics,
in both the GK and disk frameworks, as well as in the ``intermediate''
annulus framework, the massless theory admits only 't~Hooft vertices
that preserve $\U(1)_A$, just as they preserve
$\SU(N)_L \times \SU(N)_R \times \U(1)_B$.
Under these circumstances, unless the low energy theory violates some
fundamental properties, we expect that the existence of
the singlet pseudoscalar NGB cannot be avoided
when chiral symmetry breaking takes place,
just like the existence of the familiar massless pions cannot be avoided.

\section{\label{conc} Discussion}
In this paper we studied two proposals for the lattice construction
of chiral gauge theories \cite{GK,Kd,KS} having in common that
the underlying fermion system is five-dimensional.  We found that both
formulations have a superfluous conserved current which is gauge invariant.
If the target 4D theory is a chiral gauge theory,
the superfluous conserved current is the fermion number current.
If the target theory is vector-like, it is the singlet axial current.
Correspondingly, the symmetry of the vector-like theory
enlarges to $\U(N_f)_L\times\U(N_f)_R$, instead of
$\SU(N_f)_L\times\SU(N_f)_R\times \U(1)_B$ as would be expected.
In all cases, the underlying reason is that the fermion number
of each 5D field is separately conserved.

When the target 4D theory is QCD with a small enough $N_f$
to enable chiral symmetry breaking,
we concluded that the spectrum will contain a superfluous
Nambu-Goldstone boson, which is a singlet pseudoscalar meson.
Apart from chiral symmetry breaking,
we used only translation and Lorentz invariance.

The standard DWF formulation of QCD, which is also five-dimensional,
is equivalent to a purely 4D formulation in which the effective Dirac operator
satisfies the Ginsparg--Wilson relation \cite{GW}.  This raises the question
about a possible connection between the proposals of Refs.~\cite{GK,Kd},
and L\"uscher's approach to the construction of lattice chiral gauge theories
\cite{MLabelian,MLnonabelian,MLpertth}.
In the latter approach, the lattice theory is gauge invariant
provided that the fermion integration measure can be properly defined,
for which a necessary condition is that the fermion spectrum will have
no gauge anomaly.  In comparison with Refs.~\cite{GK,Kd}, a key difference
is that the total fermion number is violated in L\"uscher's approach.
This behavior is a direct consequence of the way the fermion
integration measure is defined in this approach \cite{MLabelian}.

Another approach to the construction of lattice chiral gauge theories
is the gauge fixing approach \cite{gfPT,FNV,nachgt}.
In this approach, the lattice theory
is not gauge invariant, and gauge invariance is restored only
in the continuum limit.  Like in the proposals of Refs.~\cite{GK,Kd},
in the gauge-fixing approach there is also a U(1) symmetry associated
with the total fermion number.  However, in this case the conserved current
is not gauge invariant; conversely,
the gauge invariant current is not conserved \cite{gfPT,DM}.
The U(1) fermion number symmetry
is always spontaneously broken, but the associated Nambu--Goldstone boson
does not belong to the gauge invariant physical spectrum \cite{FNV}.

We comment that the so-called ``Symmetric Mass Generation'' (SMG)
approach\footnote{
  For a review of SMG, and its relation to lattice chiral gauge theories,
  see Ref.~\cite{SMGreview}.}
to the construction of lattice chiral gauge theories does not have
a similar U(1) issue.  While gauge invariance is always maintained,
the U(1) issue is avoided.  The reason is that the multi-fermion and/or Yukawa
interactions introduced into the lattice action are designed to
break explicitly any symmetry not present in the target continuum theory,
including the fermion number symmetry.
Nevertheless this approach may fail for various dynamical reasons.
Recently, we pointed out that the SMG approach has a potential issue
with unwanted propagator zeros that could take the form of ghost states
in the continuum theory.  For details, as well as references
to the original literature, see Ref.~\cite{zero}.

It is an open question whether or not the proposals of Refs.~\cite{GK,Kd}
can lead to a consistent 4D quantum field theory in the continuum limit.
If they do, this would signal a breakdown of universality,
in the following sense.  Considering once again the QCD example,
the reason is that the new universality class would contain
a singlet pseudoscalar Nambu--Goldstone boson in the physical spectrum
if chiral symmetry breaking takes place, whereas the standard formulation
of QCD does not have such a Nambu--Goldstone boson.  If chiral
symmetry breaking does not take place, and/or the 4D theory is not
translation or Lorentz invariant,
this would also mean that the theory is different from
QCD as obtained in the continuum limit of one of the standard lattice
regularizations.
A similar breakdown of universality would happen in the case of
a chiral gauge theory, because in the proposals of Refs.~\cite{GK,Kd}
there exists a conserved and gauge invariant fermion number current,
while L\"uscher's approach does not have such a current.

\vspace{3ex}
\noindent
{\bf Acknowledgments.\ }
We thank David Kaplan and Srimoyee Sen for fruitful discussions.
This material is based upon work supported by the U.S. Department of
Energy, Office of Science, Office of High Energy Physics,
Office of Basic Energy Sciences Energy Frontier Research Centers program
under Award Number DE-SC0013682 (MG).
YS is supported by the Israel Science Foundation under grant no.~1429/21.

\appendix
\section{\label{flow} Lattice gradient flows}
In both the GK framework and the disk framework,
a 4D gauge field must be extended into 5D while preserving
the 4D gauge invariance.  Naively, one may aim for a 5D extension
of the gauge field that is smooth everywhere.  In fact,
whether or not this is possible, it is not desirable.

Let us start with the disk framework.  In order to see why
smoothness everywhere is not always possible, assume that
the initial 4D gauge field is an instanton field. Now consider
the resulting 5D gauge field in the vicinity of a cylinder defined by a circle
of radius $0<r<R$ inside the disk, together with the three
transverse directions.   If the 5D extension of the gauge field
is smooth in the vicinity of every cylinder, then the
4D topological charge on every such cylinder will be $Q=1$,
same as for the initial instanton field.
But this implies that at $r=0$ we will necessarily generate a singularity!

In the case of the GK framework,
it is possible in principle to use a continuum flow
that preserves the topological charge, without creating a singularity
anywhere in the 5D gauge field.  However, this would imply that the
Weyl field on the far wall will couple to a gauge field with the
same topological charge as the initial gauge field \cite{GK}.
The outcome will be that, instead of the (conjectured) bulk zero modes
we discussed in Sec.~\ref{tHooft}, there will be fermion zero modes
associated with the Weyl field on the far wall.  These zero modes
will replace of the bulk zero modes in the total 't~Hooft vertex
(see Eq.~(\ref{1ftot})).  However, the far-wall zero modes represent
a non-local modification of the target 4D theory, something we would like
to avoid.  Either way, the 't~Hooft vertices will preserve
the individual fermion numbers of each 5D field.

Following Refs.~\cite{GK,Kd}, in this appendix we define the 5D gauge field
via gradient flow, which automatically preserves the 4D gauge invariance.
We will introduce a family of lattice gradient flows, under which
the size of any instanton present in the initial gauge field
is expected to decrease monotonically.  Eventually, the instanton's size
becomes $O(a)$, and the (now lattice-size) instanton disappears,
leaving behind a {\em dislocation} in the 5D gauge field.

We first briefly introduce gradient flow \cite{MLGF}.
The general form of the flow equation in the continuum is
\begin{equation}
\label{GF}
\frac{\partial B_\m}{\partial t} = - g^2\, \frac{\d S}{\d B_\m} \ .
\end{equation}
Here $t\ge 0$ is the flow time parameter, and $B_\m$ is the flowed gauge field,
which is subject to the boundary condition $B_\m=A_\m$ at $t=0$, where $A_\m$
is the dynamical gauge field.  In the slab geometry, $t$ is identified
with the (cartesian) fifth coordinate $s$, while in the disk geometry
it is identified with the radial coordinate $R-r$.

The right-hand side of the flow equation is the functional derivative
of some continuum action $S$ with respect to the (flowed) gauge field.
Provided that the action $S$ that generates the flow is gauge invariant,
the flow equation is gauge covariant.
If we use the standard continuum gauge action, then the right-hand side
of the flow equation is proportional to the Yang-Mills equation of motion,
and thus all classical solutions are unmodified by the flow.
This includes in particular topologically non-trivial solutions,
such as instantons.

On the lattice, the action that generates the flow\footnote{
  For the lattice version of the flow equation, see Ref.~\cite{MLGF}.
}
can be expanded as a power series in the (squared) lattice spacing $a^2$,
much like in the Symanzik improvement program
\cite{Szik,PW,WW,LW1,LW2}.
Here we will limit ourselves to the two simplest lattice gauge actions.
The first is the single plaquette (Wilson) action
\begin{equation}
\label{Splaq}
S_W = \frac{1}{g_0^2} \sum_{x,\m\n} \Re\tr(1-P_{x,\m\n}) \ ,
\end{equation}
where $P_{x,\m\n}$ is the oriented product of link variables
around the plaquette, and  $g_0$ is the bare lattice gauge coupling.
The second is the rectangle action
\begin{equation}
\label{Srect}
S_r = \frac{2}{g_0^2} \sum_{x,\m\n} \Re\tr(1-P_{x,\m\n}^{\rm rect}) \ ,
\end{equation}
where $P_{x,\m\n}^{\rm rect}$ is the oriented product of links around the
rectangle, with $\m$ ($\n$) corresponding to the short (long) side
of the rectangle.\footnote{
Equation~(\ref{Srect}) uses the normalization of the rectangle term
most common in the literature.
}
Expanding to $O(a^2)$, these actions are given by \cite{PW,WW}
\begin{subequations}
\label{Sexp}
\begin{eqnarray}
\label{SWS3}
S_W &=& \frac{1}{g_0^2} \int d^4x
\left(\frac{1}{2}\,\tr G^2 - \frac{1}{12}\,a^2 S_3 + \cdots\right) \ ,
\\
\label{SrS3}
S_r/8 &=& \frac{1}{g_0^2} \int d^4x
\left(\frac{1}{2}\,\tr G^2 - \frac{5}{24}\, a^2 S_3 + \cdots\right) \ .
\end{eqnarray}
\end{subequations}
Here $G^2 = \sum_{\m\n} G_{\m\n}^2$
is the (squared) field strength.
The first term on the right-hand sides of both equations
is recognized as the familiar continuum gauge action.
$S_3$ is the dimension-6 operator
\begin{equation}
\label{S3}
S_3 = \sum_{\m\n} \tr\Big( (\cd_\m G_{\m\n})^2\Big) \ ,
\end{equation}
where $\cd_\m$ is the covariant derivative in the adjoint representation.
Notice that $S_3$ is invariant under hypercubic rotations only.
With $S_W$ and $S_r$ at hand,
the total lattice action that generates the flow will be assumed to be
\begin{equation}
\label{cpcr}
S = c_p S_W + c_r S_r \ .
\end{equation}
Requiring that the leading continuum term in the Symanzik expansion,
$\tr\, G^2$, will have its standard normalization implies
the constraint\footnote{The so-called Symanzik action has $c_p=5/3$
and $c_r=-1/12$, which removes the $S_3$ term from the action
in the tree approximation.
}
\begin{equation}
\label{cpcr1}
c_p + 8c_r = 1 \ .
\end{equation}

Next, if we substitute into Eqs.~(\ref{Sexp}) an instanton field
with size collective coordinate $\r$, we obtain \cite{Pierre}
\begin{subequations}
\label{Sinst}
\begin{eqnarray}
\label{SWinst}
S_W &=& \frac{8\p^2}{g_0^2} \Big(1 - \frac{1}{5}(a/\r)^2 + \cdots \Big) \ ,
\\
\label{Srectinst}
S_r/8 &=& \frac{8\p^2}{g_0^2} \Big(1 - \half(a/\r)^2 + \cdots \Big) \ ,
\end{eqnarray}
\end{subequations}
where $8\p^2/g_0^2$ is the classical instanton action.
We see that for both $S_W$ and $S_r$, the $O(a^2)$ term lowers the action,
and the effect becomes stronger if we decrease the instanton size $\r$.
Related, a general feature of the flow, both in the continuum and
on the lattice, is that the action $S$ that generates the flow is always
a monotonically decreasing function of the flow it generates \cite{MLGF}.

This motivates us to propose the following strategy:
{\em Use a lattice flow driven by the plaquette and rectangle terms,
  where both $c_p$ and $c_r$ are positive.}\
The advocated range corresponds to $0<c_p<1$,
while $c_r$ is determined via Eq.~(\ref{cpcr1}).
For any flow with these features, both $S_W$ and $S_r$ will decrease
in absolute value under the flow.  Moreover, in view of Eqs.~(\ref{Sinst}),
we may expect that the size of every instanton present in the initial
gauge field will shrink monotonically under the flow, until eventually
the instanton's size will become $O(a)$, and the instanton will disappear
at a dislocation.  For a recent related numerical study, see Ref.~\cite{AO}.

\vspace{5ex}


\begin{thebibliography}{99}

\bibi{GK}
D.~M.~Grabowska and D.~B.~Kaplan,
\ttl{Nonperturbative Regulator for Chiral Gauge Theories?,}
Phys. Rev. Lett. \textbf{116} (2016) no.21, 211602
[arXiv:1511.03649 [hep-lat]].

\bibi{Kd}
D.~B.~Kaplan,
\ttl{Chiral Gauge Theory at the Boundary between Topological Phases,}
Phys. Rev. Lett. \textbf{132} (2024) no.14, 141603
[arXiv:2312.01494 [hep-lat]].

\bibi{KS}
D.~B.~Kaplan and S.~Sen,
\ttl{Weyl Fermions on a Finite Lattice,}
Phys. Rev. Lett. \textbf{132} (2024) no.14, 141604
[arXiv:2312.04012 [hep-lat]].

\bibi{KDWF}
D.~B.~Kaplan,
\ttl{A Method for simulating chiral fermions on the lattice,}
Phys. Lett. B \textbf{288} (1992), 342-347
[arXiv:hep-lat/9206013 [hep-lat]].

\bibi{YSdwf}
Y.~Shamir,
\ttl{Chiral fermions from lattice boundaries,}
Nucl. Phys. B \textbf{406} (1993), 90-106
[arXiv:hep-lat/9303005 [hep-lat]].

\bibi{FS}
V.~Furman and Y.~Shamir,
\ttl{Axial symmetries in lattice QCD with Kaplan fermions,}
Nucl. Phys. B \textbf{439} (1995), 54-78
[arXiv:hep-lat/9405004 [hep-lat]].

\bibi{MLGF}
M.~L\"uscher,
\ttl{Properties and uses of the Wilson flow in lattice QCD,}
JHEP \textbf{08}, 071 (2010)
[erratum: JHEP \textbf{03}, 092 (2014)]
[arXiv:1006.4518 [hep-lat]].

\bibi{KSmit}
L.~H.~Karsten and J.~Smit,
\ttl{Lattice Fermions: Species Doubling, Chiral Invariance,
  and the Triangle Anomaly,}
Nucl. Phys. B \textbf{183} (1981), 103.

\bibi{NN}
H.~B.~Nielsen and M.~Ninomiya,
\ttl{Absence of Neutrinos on a Lattice. 1. Proof by Homotopy Theory,}
Nucl. Phys. B \textbf{185} (1981), 20
[erratum: Nucl. Phys. B \textbf{195} (1982), 541];
%
\ttl{Absence of Neutrinos on a Lattice. 2. Intuitive Topological Proof,}
Nucl. Phys. B \textbf{193} (1981), 173-194.

\bibi{YSan}
Y.~Shamir,
\ttl{Anomalies and chiral defects fermions,}
Nucl. Phys. B \textbf{417} (1994), 167-180
[arXiv:hep-lat/9310006 [hep-lat]].

\bibi{HK}
Y.~Hamada and H.~Kawai,
\ttl{Axial U(1) current in Grabowska and Kaplan\textquoteright{}s formulation,}
PTEP \textbf{2017} (2017) no.6, 063B09
[arXiv:1705.01317 [hep-lat]].

\bibi{hooft}
G.~'t Hooft,
\ttl{How Instantons Solve the U(1) Problem,}
Phys. Rept. \textbf{142} (1986), 357-387

\bibi{SC} S.\ Coleman, {\it The Uses of Instantons}, in
{\it Aspects of Symmetry}, Cambridge University Press, 1985.

\bibi{AF}
S.~Aoki and H.~Fukaya,
\ttl{Curved domain-wall fermion and its anomaly inflow,}
PTEP \textbf{2023} (2023) no.3, 033B05
[arXiv:2212.11583 [hep-lat]].

\bibi{AFK}
S.~Aoki, H.~Fukaya and N.~Kan,
\ttl{A lattice formulation of Weyl fermions on a single curved surface,}
PTEP \textbf{2024} (2024) no.4, 043B05
[arXiv:2402.09774 [hep-lat]].

\bibi{lcl1}
M.~Golterman and Y.~Shamir,
\ttl{Localization in lattice QCD,}
Phys. Rev. D \textbf{68} (2003), 074501
[arXiv:hep-lat/0306002 [hep-lat]].

\bibi{lcl2}
M.~Golterman, Y.~Shamir and B.~Svetitsky,
\ttl{Mobility edge in lattice QCD,}
Phys. Rev. D \textbf{71} (2005), 071502
[arXiv:hep-lat/0407021 [hep-lat]].

\bibi{lcl3}
M.~Golterman, Y.~Shamir and B.~Svetitsky,
\ttl{Localization properties of lattice fermions with
  plaquette and improved gauge actions,}
Phys. Rev. D \textbf{72} (2005), 034501
[arXiv:hep-lat/0503037 [hep-lat]].

\bibi{GW}
P.~H.~Ginsparg and K.~G.~Wilson,
\ttl{A Remnant of Chiral Symmetry on the Lattice,}
Phys. Rev. D \textbf{25} (1982), 2649

\bibi{MLabelian}
M.~L\"uscher,
\ttl{Abelian chiral gauge theories on the lattice with exact gauge invariance,}
Nucl. Phys. B \textbf{549} (1999), 295-334
[arXiv:hep-lat/9811032 [hep-lat]].

\bibi{MLnonabelian}
M.~L\"uscher,
\ttl{Weyl fermions on the lattice and the nonAbelian gauge anomaly,}
Nucl. Phys. B \textbf{568} (2000), 162-179
[arXiv:hep-lat/9904009 [hep-lat]].

\bibi{MLpertth}
M.~L\"uscher,
\ttl{Lattice regularization of chiral gauge theories to all orders of perturbation theory},
JHEP \textbf{06}, 028 (2000)
[arXiv:hep-lat/0006014 [hep-lat]].

\bibi{gfPT}
W.~Bock, M.~F.~L.~Golterman and Y.~Shamir,
\ttl{Chiral fermions on the lattice through gauge fixing: Perturbation theory,}
Phys. Rev. D \textbf{58} (1998), 034501
[arXiv:hep-lat/9801018 [hep-lat]].

\bibi{FNV}
M.~Golterman and Y.~Shamir,
\ttl{Fermion number violation in regularizations that preserve
  fermion number symmetry,}
Phys. Rev. D \textbf{67} (2003), 014501
[arXiv:hep-th/0202162 [hep-th]].

\bibi{nachgt}
M.~Golterman and Y.~Shamir,
\ttl{SU(N) chiral gauge theories on the lattice,}
Phys. Rev. D \textbf{70} (2004), 094506
[arXiv:hep-lat/0404011 [hep-lat]].

\bibi{DM}
M.~J.~Dugan and A.~V.~Manohar,
\ttl{Lattice chiral fermions and flavor anomalies,}
Phys. Lett. B \textbf{265} (1991), 137-140

\bibi{SMGreview}
J.~Wang and Y.~Z.~You,
\ttl{Symmetric Mass Generation,}
Symmetry \textbf{14}, no.7, 1475 (2022)
[arXiv:2204.14271 [cond-mat.str-el]].

\bibi{zero}
M.~Golterman and Y.~Shamir,
\ttl{Propagator Zeros and Lattice Chiral Gauge Theories,}
Phys. Rev. Lett. \textbf{132} (2024) no.8, 081903
[arXiv:2311.12790 [hep-lat]].

\bibi{Szik}
K.~Symanzik,
\ttl{Continuum Limit and Improved Action in Lattice Theories.
  1. Principles and $\varphi^4$ Theory,}
Nucl. Phys. B \textbf{226} (1983), 187-204;
\ttl{Continuum Limit and Improved Action in Lattice Theories.
  2. O(N) Nonlinear Sigma Model in Perturbation Theory,}
Nucl. Phys. B \textbf{226} (1983), 205-227

\bibi{PW}
P.~Weisz,
\ttl{Continuum Limit Improved Lattice Action for Pure Yang-Mills Theory. 1.,}
Nucl. Phys. B \textbf{212} (1983), 1-17

\bibi{WW}
P.~Weisz and R.~Wohlert,
\ttl{Continuum Limit Improved Lattice Action for Pure Yang-Mills Theory. 2.,}
Nucl. Phys. B \textbf{236} (1984), 397
[erratum: Nucl. Phys. B \textbf{247} (1984), 544]

\bibi{LW1}
M.~L\"uscher and P.~Weisz,
\ttl{On-shell improved lattice gauge theories,}
Commun. Math. Phys. \textbf{97} (1985) 59
[erratum: Commun. Math. Phys. \textbf{98} (1985), 433]

\bibi{LW2}
M.~L\"uscher and P.~Weisz,
\ttl{Computation of the Action for On-Shell Improved Lattice Gauge Theories
  at Weak Coupling,}
Phys. Lett. B \textbf{158} (1985), 250-254

\bibi{Pierre}
M.~Garcia Perez, A.~Gonzalez-Arroyo, J.~R.~Snippe and P.~van Baal,
\ttl{Instantons from over - improved cooling,}
Nucl. Phys. B \textbf{413} (1994), 535-552
[arXiv:hep-lat/9309009 [hep-lat]].

\bibi{AO}
A.~Hasenfratz and O.~Witzel,
\ttl{Dislocations under gradient flow and their effect
  on the renormalized coupling,}
Phys. Rev. D \textbf{103} (2021) no.3, 034505
[arXiv:2004.00758 [hep-lat]].

\end{thebibliography}
\end{document}